# Studiu de caz privind utilizarea modelelor IEC 61499 în controlul holonic de nivel înalt

Valentin VLAD, Adrian GRAUR, Cristina Elena TURCU, Calin CIUFUDEAN
*Ştefan cel Mare University of Suceava*
*{vladv | adriang | cristina | calin}@eed.usv.ro*

*Abstract* — În cadrul acestei lucrări sunt explorate soluții de aplicare a specificațiilor IEC 61499 în modelarea şi implementarea controlului de nivel înalt al holonilor. Conceptele IEC 61499 au fost utilizate ca o tehnologie suport pentru încapsularea controlului inteligent al holonilor, modelarea comunicației inter-holonice şi configurarea dinamică a holonilor. Modelele şi teoriile propuse au fost validate în cadrul unui sistem de control holonic dezvoltat în jurul unei celule de asamblare.

## I. INTRODUCERE

Sistemele de fabricație holonice îşi au originea în observațiile scriitorului Arthur Koestler privind modul în care sunt construite sistemele biologice şi organizațiile sociale. Koestler introduce în 1967, în cadrul lucrării „The Ghost in the Machine", termenul de *holon*, ca o combinație a două cuvinte din limba greacă: ,holos', cu semnificația de *întreg*, şi ,on', cu semnificația de *particulă*, în scopul de a descrie faptul că în cadrul unui sistem complex, elementele constituente prezintă atât un comportament de *întreg*, care poate fi divizat în subcomponente, cât şi un comportament de *componentă*, care face parte dintr-un întreg mai mare. El introduce de asemenea noțiunea de ,holarhie', definită ca o ierarhie de holoni, constituită în mod dinamic.

Aplicabilitatea conceptelor holonice în domeniul sistemele de fabricație a fost menționată pentru prima dată de către Suda, în 1990 [7] şi avea în vedere dezvoltarea unei noi generații de sisteme de control a fabricației, care să fie caracterizate de modularitate, autonomie, cooperare, distribuție şi un mod de organizare similar cu cel al sistemelor din lumea vie.

Pentru implementarea controlului holonic au fost adoptate tehnologii din domeniul sistemelor multi-agent (pentru nivelul decizional al holonilor), împreună cu tehnici de programare bazate pe blocuri funcționale (cum sunt cele definite prin IEC 61499 [9] sau IEC 61131 [4]), pentru controlul de timp real al dispozitivelor fizice. Mai mult, anumite publicații, cum ar fi [2], [3] propun utilizarea modelelor IEC 61499 atât pentru controlul „low-level" al holonilor, cât şi pentru încapsularea modulelor care compun controlul inteligent al acestora. O astfel de soluție este explorată şi în cadrul acestei lucrări, prin definirea de interfețe de blocuri funcționale pentru controlul inteligent, dezvoltarea de soluții de comunicație inter-holonice modelate prin blocuri funcționale şi utilizarea tehnicilor de reconfigurare definite de IEC 61499 pentru crearea şi eliminarea dinamică de holoni în/din cadrul unui sistem holonic.

## II. NOȚIUNI ELEMENTARE PRIVIND SPECIFICAȚIILE IEC 61499

IEC 61499 defineşte o arhitectură deschisă (eng. *open architecture*), modulară, destinată următoarei generații de sisteme din domeniul controlului distribuit şi automatizării. Standardul încorporează tehnologii software avansate cum ar fi încapsularea funcționalității, design modular, execuție controlată de evenimente şi distribuție [9], şi a fost adaptat încă din faza de proiectare pentru a fi utilizat în dezvoltarea sistemelor de control holonice [1].

Elementul de bază al arhitecturii IEC 61499 îl reprezintă *blocul funcțional*, definit ca o unitate compozițională cu o interfață specifică şi un set de elemente de funcționalitate, care pot fi implementate atât prin software, cât şi prin hardware. IEC 61499 extinde conceptul de *blocuri funcționale* definit de IEC 61131, prin integrarea acestora într-un cadru global de dezvoltare a sistemelor distribuite.

Interfața unui bloc funcțional IEC 61499 (Fig. 1) este definită ca o listă de intrări de evenimente, intrări de date, ieşiri de evenimente şi ieşiri de date, la care se adaugă intrări de tip „socket" şi ieşiri de tip „fişe", pentru blocurile de tip *adaptor*. Reprezentarea grafică a unui bloc funcțional poate fi văzută ca fiind compusă din două componente, denumite „cap" (partea superioară) şi „trup" (partea inferioară). *Capul* conține intrările şi ieşirile de evenimente, în timp ce intrările şi ieşirile de date sunt conectate la *trupul* blocului funcțional.

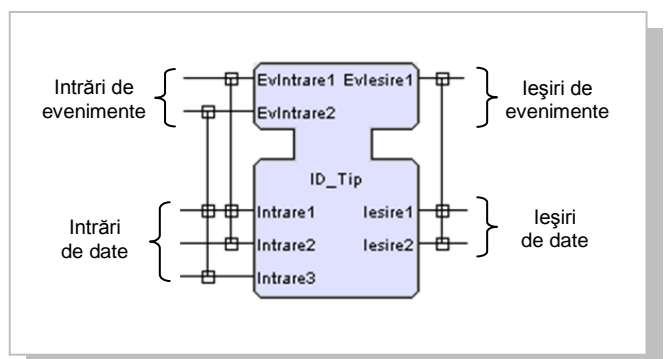

Fig. 1. Interfața unui bloc funcțional

Activarea unui bloc funcţional este realizată prin intermediul evenimentelor conectate la acesta, de unde şi expresia de „execuţie controlată de evenimente", care reprezintă o caracteristică de bază a standardului. Evenimentele permit sincronizarea şi interacţiunea dintre mecanismele de control din cadrul unei reţele de blocuri funcţionale. Trimiterea datelor între două blocuri funcţionale este însoţită totdeauna de generarea unui eveniment de către blocul emiţător, care asigură sincronizarea şi citirea corectă a datelor la receptor [9].

Alte elemente cheie definite în cadrul standardului sunt *aplicaţia*, *resursa*, *dispozitivul* şi *sistemul*. O *aplicaţie* este văzută ca o reţea de blocuri funcţionale interconectate între ele, care defineşte în mod complet, dar abstract, funcţionalitatea sistemului, fără a lua în considerare modul în care va fi distribuită la nivelul dispozitivelor de procesare. Aplicaţiile pot fi structurate ierarhic prin gruparea blocurilor funcţionale care le compun în cadrul unor *subaplicaţii*. O *resursă* poate fi privită ca un container pentru o reţea de blocuri funcţionale şi corespunde unui procesor, fir de execuţie sau task (în terminologia IEC 61131-3), care execută o parte din aplicaţia distribuită. Un *dispozitiv* reprezintă o unitate de control care poate conţine zero, una sau mai multe resurse, împreună cu interfeţe de proces şi interfeţe de comunicaţie. Un *sistem* este definit ca un set de dispozitive interconectate prin reţele de comunicaţie.

Tehnologiile software deţinute de IEC 61499 permit aplicarea sa atât în dezvoltarea componentelor „low-level" ale aplicaţiilor de control, cât şi în dezvoltarea unor componente software inteligente, cum ar fi cazul unor *agenţi*. Standardul nu defineşte însă nici un protocol pentru interacţiunea dintre aceste *componente inteligente*, acest lucru rămânând subiectul unor dezvoltări ulterioare.

### III. ASPECTE PRIVIND MODELAREA ŞI IMPLEMENTAREA HOLONILOR

Arhitectura PROSA [8] identifică în cadrul unui sistem de fabricaţie holonic trei tipuri de holoni de bază: holoni resursă, holoni produs şi holoni comandă. Holonii resursă gestionează echipamentele de producţie din sistem, holonii produs sunt responsabili de planificarea tehnologică a produselor, în timp ce holonii comandă coordonează execuţia diferitelor comenzi primite de la utilizatorii sistemului.

În cadrul acestei lucrări a fost considerată pentru holonii resursă o arhitectură pe trei niveluri (Fig. 2), similară celei propuse în [1]. Un holon resursă va conţine una sau mai multe *entităţi non-holonice*, reprezentate de dispozitive mecatronice simple şi logica de control asociată, care acoperă *nivelul fizic*, respectiv cel al *logicii de control*. Entităţile non-holonice sunt coordonate de o *componentă inteligentă*, aflată pe *nivelul de coordonare*, care încapsulează controlul de nivel înalt al holonului.

Ca şi exemplu pentru acest tip de holon se poate considera o celulă de asamblare constituită dintr-un robot de asamblare şi un set de dispozitive de fixare şi transport, care conlucrează sub coordonarea unei componente software inteligente.

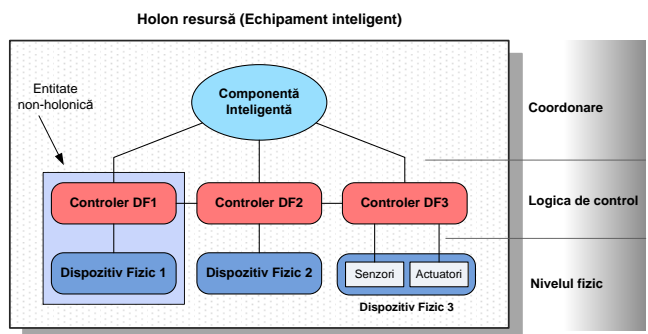

Fig. 2. Arhitectura unui holon resursă

În cadrul lucrării s-a propus dezvoltarea controlului inteligent al unui holon în forma a două componente: o componentă *conştientă*, cu rol în realizarea de raţionamente complexe şi participarea în procese de negociere, şi o componentă *subconştientă*, care să asigure gestionarea conexiunilor „de lucru" (de timp real) dintre holoni, şi rezolvarea de probleme simple ridicate de controlul fizic al acestora, prin utilizarea de soluţii predefinite.

Holonii individuali pot fi grupaţi în holoni complecşi, permiţând formarea de holarhii multi-nivel. Ca şi exemplu se poate considera cazul servirii unei comenzi primite de la un client. La primirea unei comenzi este instanţiat un *holon comandă* (holon individual), care analizează specificaţiile produsului care se doreşte a fi realizat şi decide dacă acesta este un *produs simplu* sau un *produs compus*. Un produs simplu este considerat ca fiind realizabil doar prin operaţii de prelucrare (de exemplu strunjire, frezare etc.), aplicate unei singure componente semifabricat, în timp ce un produs compus este constituit din mai multe produse simple, asamblate pe diferite niveluri (Fig. 3).

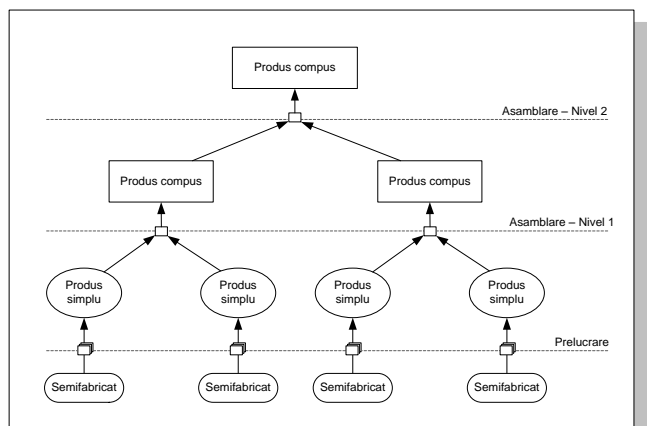

Fig. 3. Modul de realizare a produselor simple şi compuse

În cazul în care produsul comandat de client este unul simplu, holonul comandă va începe un proces de negociere cu holonii resursă din sistem, capabili să execute operaţiile produsului. Rezultatul negocierilor va consta în formarea unei „echipe" (holon complex) constituită din holonii resursă selectaţi pentru execuţia operaţiilor, şi holonul comandă, care deţine rolul de *coordonator* al echipei.

Considerând cazul unui produs compus, acesta poate fi realizat prin asamblarea mai multor componente, fie simple, fie, la rândul lor, compuse. Un prim pas în realizarea produsului ar consta în verificarea disponibilităţii componentelor. În cazul în care acestea există pe stoc, echipa de realizare a produsului va consta din holonul comandă, holonul corespunzător depozitului (care asigură livrarea componentelor), holonul sistemului de transport şi un holon resursă selectat pentru realizarea asamblării implicată de produs. În caz contrar, holonul comandă va instanţia alţi holoni comandă, de nivel inferior, care se vor ocupa de realizarea componentelor. Procesul de descompunere a activităţilor produsului va continua până la nivelul produselor simple, sau al celor aflate pe stoc, conducând la formarea de holarhii multi-nivel (Fig. 4). În cadrul acestor holarhii, planificările de pe nivelurile superioare se vor realiza funcţie de cele de pe nivelurile inferioare şi pot fi influenţate, în timpul execuţiei, de modificările acestora, datorate apariţiei unor perturbaţii. În Fig. 4 comanda clientului, de realizare a produsului A, este preluată de holonul comandă al holonului A. Acesta constată că produsul A este compus din două componente, B şi C, care nu se găsesc pe stoc, şi instanţiază, pentru realizarea acestora, doi holoni comandă de ordin inferior, care vor forma holonii B şi C. Produsul C este un produs simplu şi, ca urmare, holonul C va avea o structură plată, fiind constituit doar din holoni cu inteligenţă simplă (holoni de prelucrare, holon depozit, holon transport, holon comandă). Produsul B este însă compus din componentele D şi E, care nu sunt disponibile, ceea ce conduce la formarea sub-holonilor D şi E. Produsul D este simplu, şi ca urmare structura holonului D va avea un singur nivel (în mod similar holonului C), iar componentele produsului E (care este compus) sunt disponibile pe stoc, nefiind necesară instanţierea de noi sub-holoni.

Odată cu finalizarea unui produs, holonul comandă corespunzător acestuia va fi eliminat, iar echipa (sau holarhia) de realizare a produsului „desfiinţată".

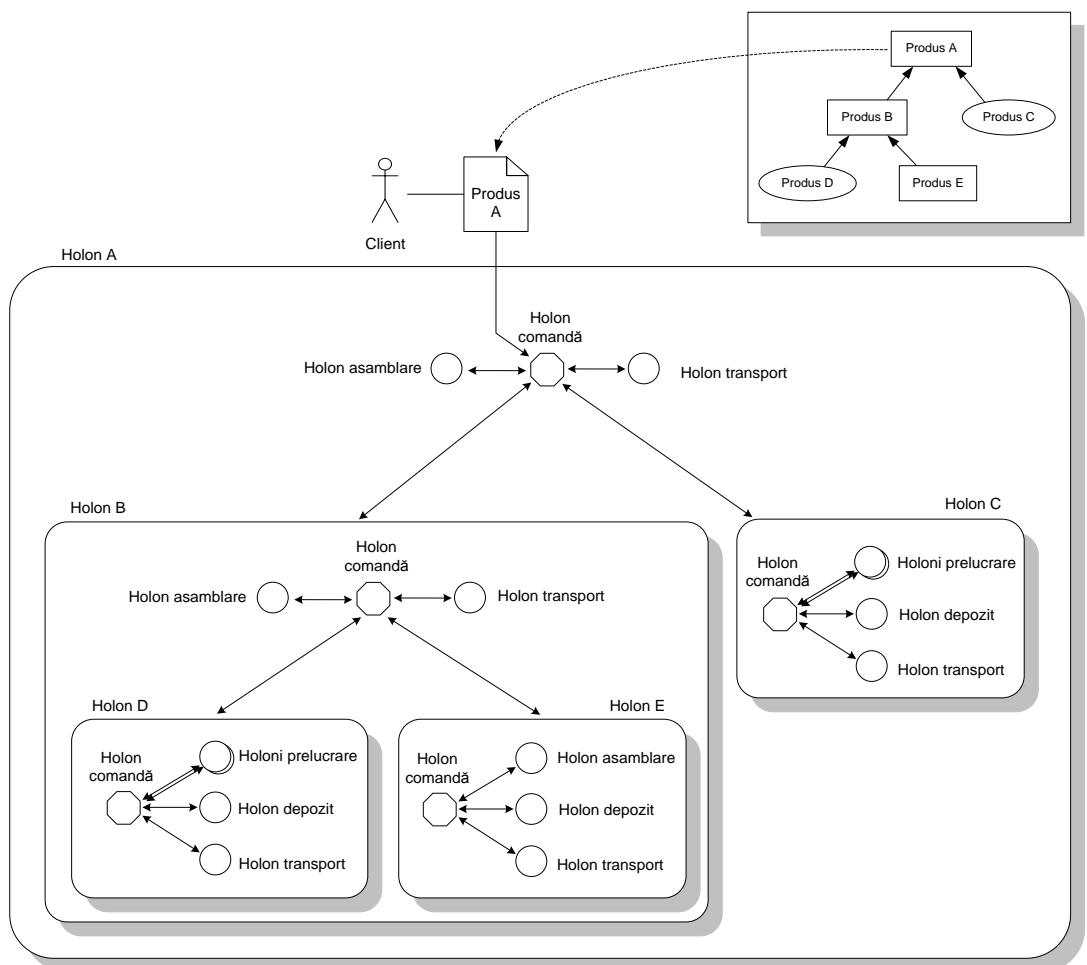

Fig. 4. Exemplu de formare a holonilor complecşi în vederea execuţiei unei comenzi

*A. Soluţii de modelare şi implementare a holonilor individuali prin blocuri funcţionale*

În cadrul acestui paragraf sunt analizate mai în detaliu structura şi soluţiile de modelare/implementare propuse pentru holonii resursă individuali, avându-se în vedere în principal partea software (nivelul de coordonare şi nivelul de control „low-level") a acestora.

Pentru implementarea unui holon individual au fost considerate necesare şase tipuri de module, reprezentate de:

- Componenta conştientă a holonului;
- Componenta subconştientă a holonului;
- Interfaţa inter-holonică;
- Controlul fizic;
- Interfaţa dintre controlul inteligent şi controlul fizic, denumită *interfaţă independentă de hardware*;
- Interfaţa om-maşină (HMI) a holonului.

Modul de interacţiune dintre aceste module este prezentat în Fig. 5.

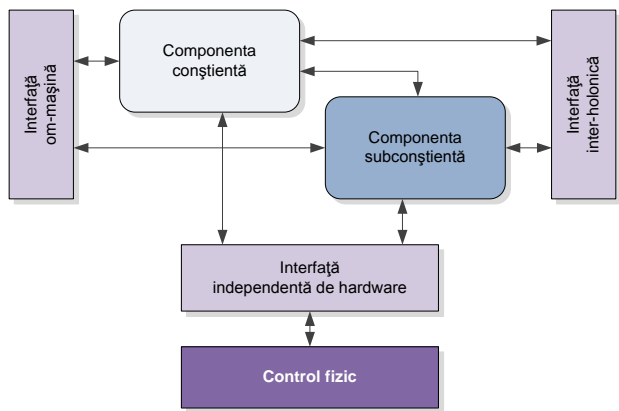

Fig. 5. Comunicaţia dintre modulele software ale unui holon individual

Cele două componente care asigură controlul inteligent al holonilor (componenta conştientă şi cea subconştientă) au fost modelate/implementate în cadrul acestei lucrări prin blocuri funcţionale IEC 61499 de interfaţare a serviciilor, cu interfeţe similare celor din Fig. 6. Aceste blocuri sunt încapsulate în cadrul unor blocuri funcţionale compuse, împreună cu un set de blocuri de comunicaţie (de tipul Publish/Subscribe), care permit comunicaţia componentelor cu celelalte module ale holonului.

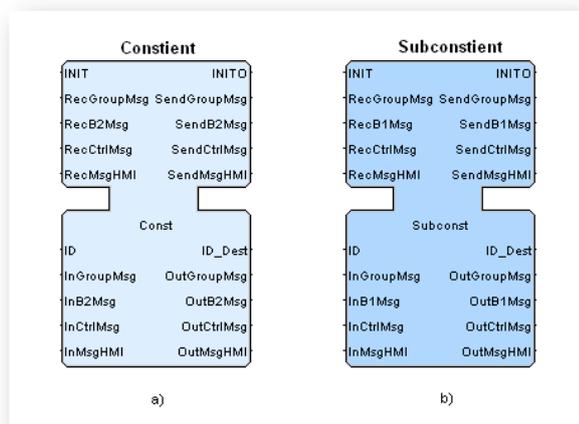

Fig. 6. Interfeţele blocurilor funcţionale pentru modelarea componentei conştiente (a) şi subconştiente (b) ale unui dispozitiv holonic

Tabelul 1 conţine o descriere a interfeţelor blocurilor funcţionale pentru implementarea componentelor controlului inteligent. Interfeţele includ intrări şi ieşiri de date (asociate cu intrări şi ieşiri de evenimente) pentru recepţia, respectiv trimiterea de mesaje de la/către modulele holonului cu care componentele sunt conectate. Această formă a interfeţelor poate fi considerată *tipică* pentru modelarea controlului inteligent, favorizând o standardizare a structurii şi modului de implementare al holonilor.

*Interfaţa inter-holonică* permite unui holon transmiterea şi recepţia de mesaje către şi de la ceilalţi holoni din cadrul sistemului. Aşa cum se poate observa şi din Fig. 7, această interfaţă a fost dezvoltată în forma a trei blocuri funcţionale, de tipul SUBSCRIBE, Dispatcher şi PUBLISH. Blocul SUBSCRIBE asigură recepţia de mesaje de la ceilalţi holoni, şi livrarea lor către blocul Dispatcher. Blocul Dispatcher va extrage antetul mesajului şi, în baza informaţiilor conţinute de acesta, va dirija mesajul către componenta conştientă sau către componenta subconştientă a holonului. Blocul PUBLISH permite atât componentei conştiente, cât şi celei subconştiente, transmiterea de mesaje către alţi holoni.

Tabelul 1. Descrierea intrărilor/ieşirilor de evenimente şi date ale componentelor controlului inteligent

| Intrări/ieşiri de evenimente/date | Descriere |
|---|---|
| RecGroupMsg | Eveniment care însoţeşte un mesaj de la un alt holon |
| RecB2Msg | Eveniment care însoţeşte un mesaj de la componenta subconştientă |
| RecB1Msg | Eveniment care însoţeşte un mesaj de la componenta conştientă |
| RecCtrlMsg | Eveniment care însoţeşte un mesaj de la logica de control |
| RecMsgHMI | Eveniment care însoţeşte un mesaj de la interfaţa HMI |
| ID | ID-ul holonului (identic pentru ambele componente) |
| InGroupMsg | Mesaj de la un alt holon |
| InB2Msg | Mesaj de la componenta subconştientă |
| InB1Msg | Mesaj de la componenta conştientă |
| InCtrlMsg | Mesaj de la logica de control |
| InMsgHMI | Mesaj de la interfaţa HMI |
| SendGroupMsg | Eveniment care însoţeşte un mesaj către un alt holon |
| SendB2Msg | Eveniment care însoţeşte un mesaj pentru componenta subconştientă |
| SendB1Msg | Eveniment care însoţeşte un mesaj pentru componenta conştientă |
| SendCtrlMsg | Eveniment care însoţeşte un mesaj pentru logica de control |
| SendMsgHMI | Eveniment care însoţeşte un mesaj pentru interfaţa HMI |
| ID_Dest | ID-ul holonului destinaţie |
| OutGroupMsg | Mesaj către un alt holon |
| OutB2Msg | Mesaj către componenta subconştientă |
| OutB1Msg | Mesaj către componenta conştientă |
| OutCtrlMsg | Mesaj pentru logica de control |
| OutMsgHMI | Mesaj pentru interfaţa HMI |

Pentru conţinutul mesajelor vehiculate între agenţi, a fost dezvoltat şi implementat un protocol de comunicaţie de nivel înalt (care asigură suport pentru activităţi de negociere şi cooperare între holoni), bazat pe limbajul XML. În cadrul acestui protocol fiecare mesaj are forma unui element XML, în care tipul mesajului este dat de numele elementului, iar datele mesajului sunt încapsulate în atributele şi conţinutul elementului.

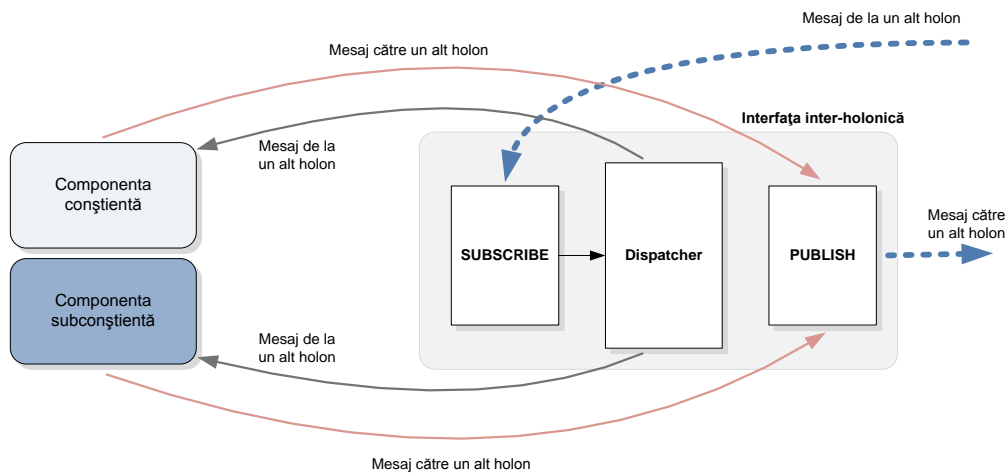

Fig. 7. Comunicaţia dintre componentele controlului inteligent şi interfaţa inter-holonică

Fig. 8 conţine două exemple de mesaje transmise între holoni, în vederea negocierii execuţiei unei operaţii. Primul mesaj (de tipul GetBidForOp) reprezintă o *cerere de ofertă* pentru execuţia unei operaţii cu codul Op_30, care poate să înceapă cel mai devreme la timpul 1308574904, (reprezentând numărul de secunde scurse de la Epoch - 1 Ianuarie 1970), echivalent cu data de 20/06/2011, ora 16:01:44. Al doilea exemplu reprezintă răspunsul unui holon resursă la un mesaj de tipul GetBidForOp, indicând faptul că execuţia operaţiei cu codul Op_30 poate să înceapă cel mai devreme la timpul 1308574950 (adică 20/06/2011 16:02:30) şi durează 50 de secunde.

Accesul componentelor controlului inteligent la controlul echipamentelor fizice se realizează prin intermediul unor *interfeţe independente de hardware*, implementate în forma unor blocuri funcţionale de interfaţare a serviciilor. Atunci când un holon deţine dispozitive fizice controlate în formă clasică, soluţia adoptată are în vedere dezvoltarea de interfeţe independente de hardware pentru fiecare controler (PLC, controler CNC etc.) asociat acestor dispozitive. În acest caz interfeţele independente de hardware au rolul de a abstractiza comenzile specifice unui anumit dispozitiv, punând la dispoziţia controlului inteligent un set de servicii şi comenzi de nivel înalt. În cazul în care logica de control are forma unui controler (distribuit) IEC 61499, un holon va conţine o singură interfaţă independentă de hardware, inclusă în acel controler.

```
<GetBidForOp ID="15" OpID="Op_30" MinStartTime="1298574904"
    Sender="225.0.0.1:2101" />
```

```
<RspBidForOp ID="15" OpID="Op_30" StartTime="1298574950"
    ExecTime="50" Sender="225.0.0.1:3001" />
```

Fig. 8. Exemplu de mesaje transmise între holoni

*Interfaţa om-maşină* a unui holon permite utilizatorilor umani monitorizarea stării holonului şi intervenţia în funcţionarea acestuia. Ca şi soluţie de implementare, în cadrul acestei lucrări s-a ales dezvoltarea acestui modul ca o reţea de blocuri funcţionale IEC 61499 de interfaţare a serviciilor, dintre care unul are rolul de a asigura comunicaţia interfeţei cu componentele controlului inteligent, iar celelalte permit afişarea a diferite controale grafice (butoane, casete text, diagrame etc.).

Ca urmare, fiecare modul din componenţa unui holon individual, cu excepţia modulelor de control clasice, poate fi modelat/implementat prin blocuri funcţionale IEC 61499, şi mapat în cadrul unui dispozitiv IEC 61499, cum este ilustrat în Fig. 9.

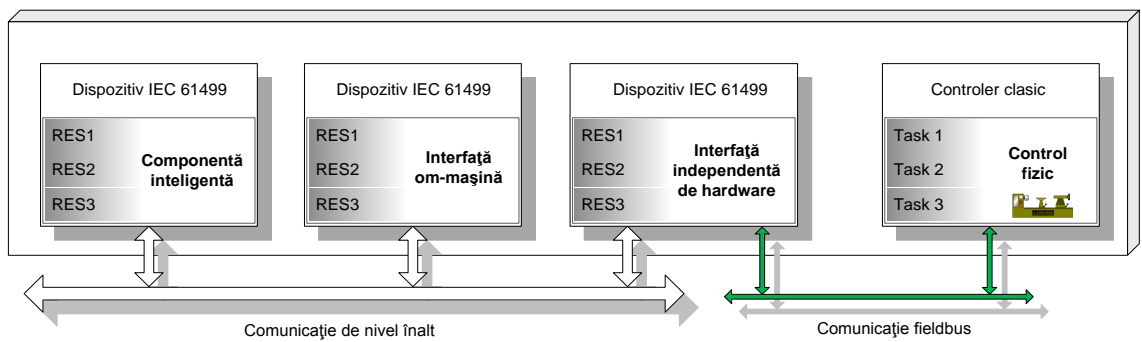

Fig. 9. Maparea componentelor unui holon individual în dispozitive IEC 61499

## IV. APLICAŢIE PRACTICĂ DE IMPLEMENTARE A UNUI SISTEM DE CONTROL HOLONIC

În cadrul acestui capitol este prezentată o aplicaţie practică prin care au fost validate soluţiile de adoptare a conceptelor IEC 61499 în controlul holonic, prezentate în capitolul anterior. Sistemul holonic de control a fost dezvoltat în jurul unui echipament de producţie controlat în formă clasică, reprezentat de o celulă de asamblare robotizată din inventarul Facultăţii de Inginerie Electrică şi Ştiinţa Calculatoarelor, Universitatea Ştefan cel Mare din Suceava. O motivaţie în plus pentru alegerea celulei în cadrul acestui studiu de caz o reprezintă gradul ridicat de utilizare a unor astfel de sisteme în domeniul industrial.

### A. Celula de asamblare

După cum se poate observa şi din Fig. 10, celula de asamblare este constituită în principal dintr-un robot de asamblare SCARA (model Sony SRX-611) şi un set de dispozitive de transport şi fixare, acţionate electric sau pneumatic.

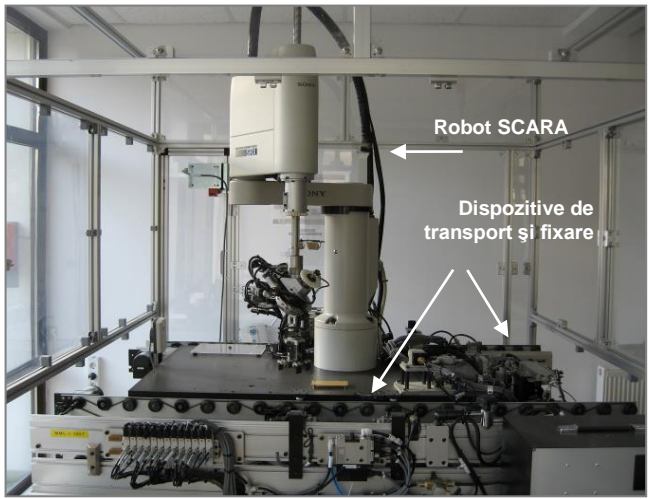

Fig. 10. Celula de asamblare

Din punct de vedere al controlului celula este echipată cu două controlere (Fig. 11), unul reprezentat de controlerul robotului (SRX-C61) şi celălalt de un PLC (Omron C200HX) în care rulează o aplicaţe pentru controlul dispozitivelor de transport şi fixare. Cele două controlere sunt conectate şi comunică între ele prin intermediul unor intrări şi ieşiri digitale. De asemenea, ambele controlere dispun de porturi de comunicaţie RS232, care le permit o comunicaţie punct la punct cu un dispozitiv extern celulei, pe baza unui anumit protocol.

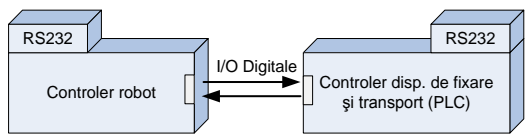

Fig. 11. Controlerele celulei de asamblare

### B. Modelul holonic al celulei de asamblare

În Fig. 12 este prezentată arhitectura software a holonului corespunzător celulei de asamblare. Aceasta include aplicaţiile din cele două controlere ale celulei, interfeţele independente de hardware ale controlului fizic şi componentele controlului inteligent.

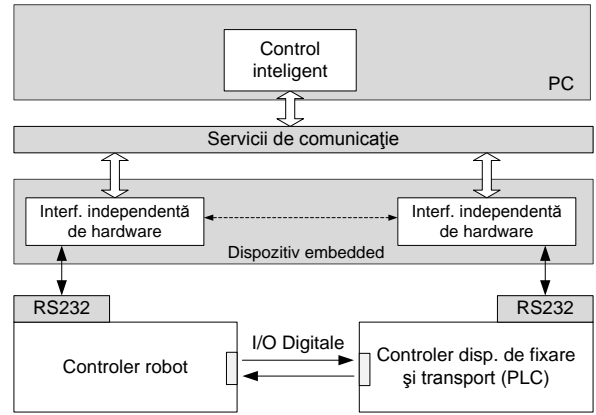

Fig. 12. Arhitectura software a holonului celulei de asamblare

În implementarea actuală cele două module pentru interfeţele independente de hardware rulează pe un dispozitiv embedded, iar "controlul inteligent" rulează pe un sistem de calcul de tip PC, cu resurse de procesare superioare. Comunicaţia dintre controlul inteligent şi interfeţele independente de hardware utilizează un mecanism de tipul Publish/Subscribe bidirecţional (protocol UDP/IP), care permite ambelor capete ale comunicaţiei să iniţieze transferuri de date.

Dispozitivul embedded poate fi privit ca un controler adiţional ataşat celulei, care asigură atât interfaţarea fizică a comunicaţiei (RS232 - Ethernet), cât şi pe cea logică, punând la dispoziţia controlului inteligent un set de servicii şi comenzi de nivel înalt.

### C. Modelul operaţional al sistemului holonic de asamblare

Fig. 13 prezintă modelul operaţional simplificat al sistemului holonic dezvoltat în jurul celulei de asamblare.

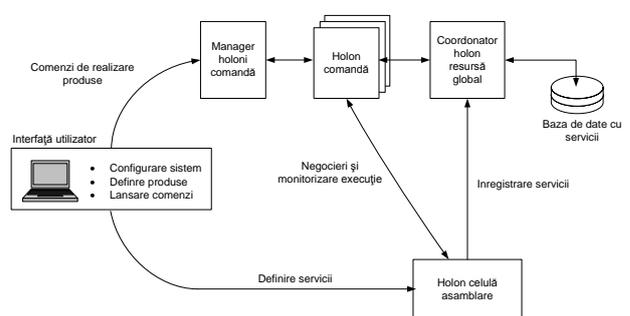

Fig. 13. Modelul operaţional al sistemului holonic de asamblare

Componentele modelului sunt reprezentate de o interfaţă utilizator (pentru configurarea sistemului, definirea de produse şi lansarea de comenzi), o bază de date cu serviciile oferite de sistem, managerul holonilor comandă, coordonatorul holonului resursă global, holonul celulei de asamblare şi holonii comandă, instanţiaţi funcţie de comenzile primite de la utilizatori.

*Interfaţa utilizator* permite definirea de servicii pentru celula de asamblare şi lansarea de comenzi pentru realizare de produse. În implementarea actuală un anumit serviciu este definit ca fiind compus dintr-un set de operaţii de asamblare simple, de tipul *pick and place*, care pot fi stocate în controlerul robotului.

*Managerul holonilor comandă* asigură crearea dinamică de holoni comandă, funcţie de comenzile primite de la interfaţa cu utilizatorul, şi eliminarea acestora, o dată cu finalizarea produselor care le-au fost atribuite.

*Coordonatorul holonului resursă global* asigură gestionarea bazei de date a serviciilor, prin înregistrarea de servicii în aceasta şi furnizarea, către holonii comandă, a adresei holonilor responsabili pentru un anumit serviciu.

*Holonii comandă* interacţionează, cu managerul holonilor comandă, în vederea instanţierii de holoni comandă copil, şi cu *holonul celulei de asamblare*, în vederea planificării task-urilor implicate de produse şi achiziţiei de informaţii privind starea curentă a execuţiei.

*Baza de date a serviciilor* (implementată în forma unei baze de date MySQL [5]) conţine o asociere între serviciile din sistem şi dispozitivele holonice care le pot realiza.

*D. Aspecte privind modelarea IEC 61499 şi implementarea sistemului holonic considerat*

Soluţia aleasă în cadrul acestei lucrări pentru implementarea sistemelor holonice are în vedere utilizarea specificaţiilor standardului IEC 61499 atât pentru controlul „low-level" (încorporat în controlerele dispozitivelor fizice) cât şi pentru implementarea componentelor inteligente din cadrul sistemului şi a comunicaţiei dintre ele.

Totuşi, considerând cazul integrării unor dispozitive clasice în cadrul unui sistem holonic, controlerele acestor dispozitive nu permit rularea de aplicaţii conforme cu specificaţiile IEC 61499, standardul predominant în prezent fiind IEC 61131. Ca urmare, aplicaţiile din controlerele acestor dispozitive vor fi dezvoltate în formă clasică, rămânând ca interfeţele independente de hardware împreună cu componentele inteligente din cadrul sistemului să fie dezvoltate conform IEC 61499.

Pentru implementarea aplicaţiei de control IEC 61499 a fost folosit cu precădere mediul de dezvoltare $O^3$neida Fbench [6]. Elementele hardware care compun sistemul de asamblare sunt ilustrate în Fig. 14 şi constau în principal dintr-un sistem de calcul de tip PC, un dispozitiv embedded cu suport pentru rularea de aplicaţii Java (Elsist Netmaster) şi celula de asamblare Sony.

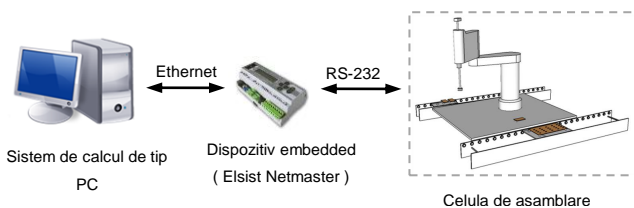

Fig. 14. Elementele hardware care compun sistemul de asamblare

Pe sistemul de calcul de tip PC rulează managerul holonilor comandă, coordonatorul holonului resursă global, holonii comandă, holonii serviciu şi controlul inteligent al celulei Sony. Pe dispozitivul embedded rulează componentele pentru interfeţele independente de hardware, iar în controlerele celulei (în controlerul robotului şi PLC-ul Omron) rulează aplicaţii pentru controlul activităţilor robotului şi ale dispozitivelor de transport şi fixare.

*D. Aplicaţia distribuită IEC 61499*

Aplicaţia de control IEC 61499 a fost dezvoltată, pe baza modelului propus, în cadrul mediului $O^3$neida Fbench, în forma unui *sistem* cu patru dispozitive: HMI, OrderHolons, Assembling_DEV şi NETMASTER (Fig. 15).

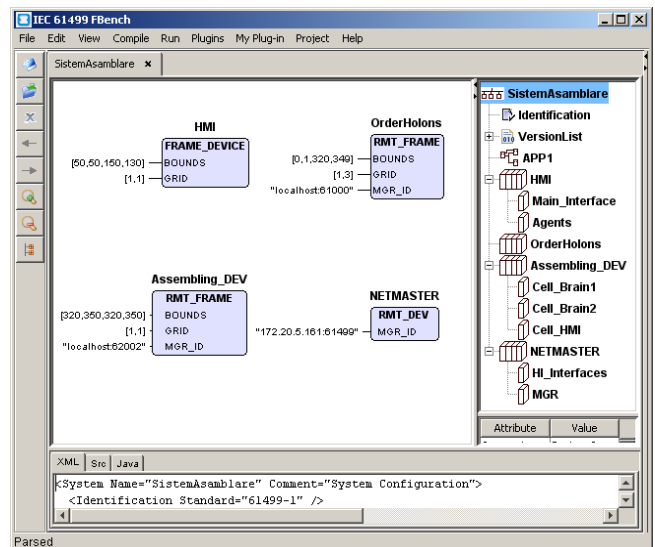

Fig. 15. Configuraţia sistem corespunzătoare aplicaţiei de control IEC 61499

Dispozitivul HMI include interfaţa utilizator a sistemului, managerul holonilor comandă şi coordonatorul holonului resursă global. Dispozitivul OrderHolons este utilizat pentru crearea/eliminarea dinamică de holoni comandă (prin comenzi de management), dispozitivul Assembling_DEV include controlul inteligent şi interfaţa HMI a celulei de asamblare, iar dispozitivul NETMASTER conţine interfeţele independente de hardware ale celor două controlere ale celulei.

*E. Scenariu de lucru*

Scenariul de lucru a fost ales aşa încât să implice utilizarea de holoni complecşi, obţinuţi prin descompu-nerea task-urilor în subtask-uri. Practic, aplicaţia are în vedere plantarea în diferite configuraţii a unor componente electronice pe nişte plăcuţe de test, şi asamblarea acestora, două câte două, în cadrul unor produse finite compuse. Fiecare configuraţie de plasare a componentelor electronice, respectiv procesul de asamblare a două plăcuţe într-un singur produs, sunt privite ca *servicii* pe care celula le poate realiza.

Componentele electronice sunt preluate dintr-o *magazie de componente*, reprezentată de o plăcuţă de test cu o dimensiune mai mare, transportată de conveierul principal al celulei (Fig. 16), pe care sunt aşezate în prealabil şi într-o configuraţie predefinită un anumit număr de componente. Plăcuţele de test pe care urmează a fi plantate componentele (denumite *plăcuţe semifabricat*) sunt transportate de conveierul secundar al celulei. Ambele conveiere conţin câte un post de lucru, în care sunt blocaţi paleţii care transportă cele două tipuri de plăcuţe, înainte de începerea procesului de asamblare.

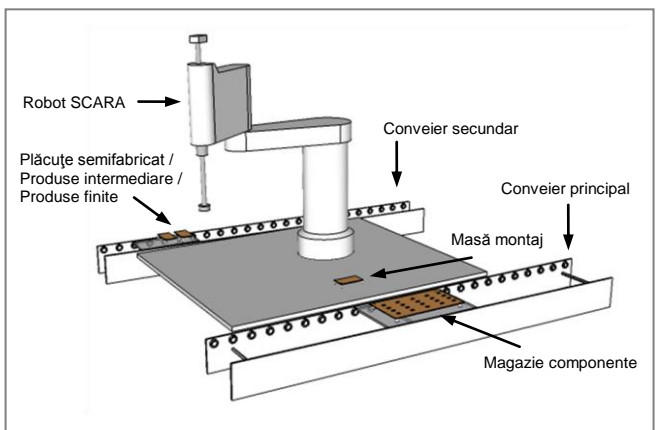

Fig. 16. Reprezentarea simplificată a celulei şi procesului de asamblare

Sistemul de asamblare poate primi comenzi pentru realizarea de *produse intermediare*, constând în plăcuţe semifabricat pe care sunt plantate componente electronice în diverse configuraţii, sau pentru realizarea de *produse finite*, formate prin asamblarea a câte două produse intermediare.

Realizarea unui produs intermediar implică următorii paşi:
- Plăcuţa semifabricat este preluată de pe paletul care o transportă şi aşezată pe masa de montaj;
- Sunt preluate componente electronice din *magazie* şi aşezate pe plăcuţa semifabricat, în configuraţia corespunzătoare serviciului cerut de holonul comandă asociat produsului;
- Produsul obţinut este aşezat înapoi pe conveierul secundar pentru a fi transportat mai departe sau pentru a fi folosit în realizarea unui produs finit.

Configuraţiile de plasare a componentelor electronice pe plăcuţele semifabricat pot fi definite în prealabil cu ajutorul unei aplicaţii speciale, şi stocate în fişiere. Un număr limitat de astfel de configuraţii pot fi stocate temporar în memoria robotului, reprezentând servicii pe care celula le poate realiza.

Realizarea unui produs finit implică existenţa a două produse intermediare pe paletul conveierului secundar, care vor fi preluate de robot şi asamblate pe masa de montaj prin îmbinarea a două perechi de conectori, plasaţi la extremităţile plăcuţelor (Fig. 17).

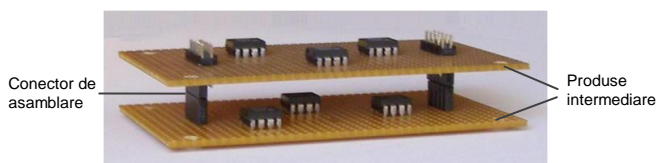

Fig. 17. Exemplu de produs finit

Alegerea unei soluţii de control holonice permite planificarea individuală a fiecărui produs şi, ca urmare, servirea de comenzi variate şi în cantităţi reduse, cu costuri de reconfigurare minime a celulei.

Informaţiile de procesare a produselor au fost structurate în forma unor documente XML, care conţin tipul produsului (simplu sau compus) şi serviciile implicate de acesta. În cadrul acestui studiu de caz produsele intermediare au fost tratate ca şi produse simple, iar produsele finite ca produse compuse. În Tabelul 2 şi Tabelul 3 este exemplificat modul de descriere a fiecăruia din aceste tipuri. Produsul ilustrat în Tabelul 2, denumit P_20, este unul simplu, şi necesită un singur serviciu, cu identificatorul S_20. Produsul P_10, prezentat în Tabelul 3, este compus din produsele intermediare P_20 şi P_21, care pot fi asamblate împreună prin serviciul cu codul S_10.

Tabelul 2 . Exemplu de document XML cu informaţii de procesare a unui produs simplu

```
<Product Name="P_20" Type="Simple">
  <Service Index="1" ServID="S_20"/>
</Product>
```

Tabelul 3. Exemplu de document XML cu informaţii de procesare a unui produs compus

```
<Product Name="P_10" Type="Composite">
  <Service Index="1" ServID="S_10" NrCmp="2"
       Cmp1="P_20" Cmp2="P_21"/>
</Product>
```

Se presupune în continuare că un utilizator lansează o comandă de realizare a unui produs de tipul P_10, care are asociat, ca informaţii de procesare, documentul XML din Tabelul 3. Ca urmare, managerul holonilor comandă va crea un holon comandă, căruia îi va transmite comanda lansată. Holonul comandă va analiza documentul XML al produsului şi va constata că este un produs compus, pentru care sunt necesare o componentă de tipul P_20 şi una de tipul P_21. În aceste condiţii holonul comandă va cere managerului holonilor comandă instanţierea a doi holoni comandă copil, care să se ocupe de realizarea componentelor. Se presupune că cele două componente au o descriere similară celei din Tabelul 2.

Considerând baza de date a serviciilor ca având forma exemplificată în Fig. 18, holonii comandă ai produselor P_20 şi P_21 vor negocia cu holonul cu adresa 225.0.0.1:3002, reprezentând holonul celulei de asamblare, pentru planificarea serviciilor (S_20 şi S_21) implicate de realizarea produselor. După finalizarea negocierilor, cei doi holoni îşi vor trimite planurile de execuţie către holonul părinte, responsabil de asamblarea produselor P_20 şi P_21 într-un singur produs (P_10). Funcţie de timpul de finalizare a celor două subcomponente, holonul produsului P_10 va negocia cu holonul celulei timpul pentru executarea serviciului S_10.

Fig. 18. Conţinutul bazei de date a serviciilor

Holonul celulei va plasa comenzile primite într-o coadă de aşteptare, şi le va executa funcţie de disponibilitatea componentelor electronice şi a plăcuţelor semifabricat.

În Fig. 19 sunt exemplificate interfeţele om-maşină ale holonilor comandă şi a holonului celulei pentru cazul unei comenzi de realizare a unui produs de tipul P_10. Interfeţele holonilor comandă permit monitorizarea gradului de finalizare al produselor, în timp ce interfaţa holonului celulei permite vizualizarea gradului de încărcare al acesteia şi a ordinii de execuţie a serviciilor. Din figură se poate observa cum mai întâi au fost planificate serviciile S_20 şi S_21 (notate cu „20" şi „21" în cadrul diagramelor Gantt), pentru realizarea produselor intermediare, şi în final serviciul S_10, pentru asamblarea acestora.

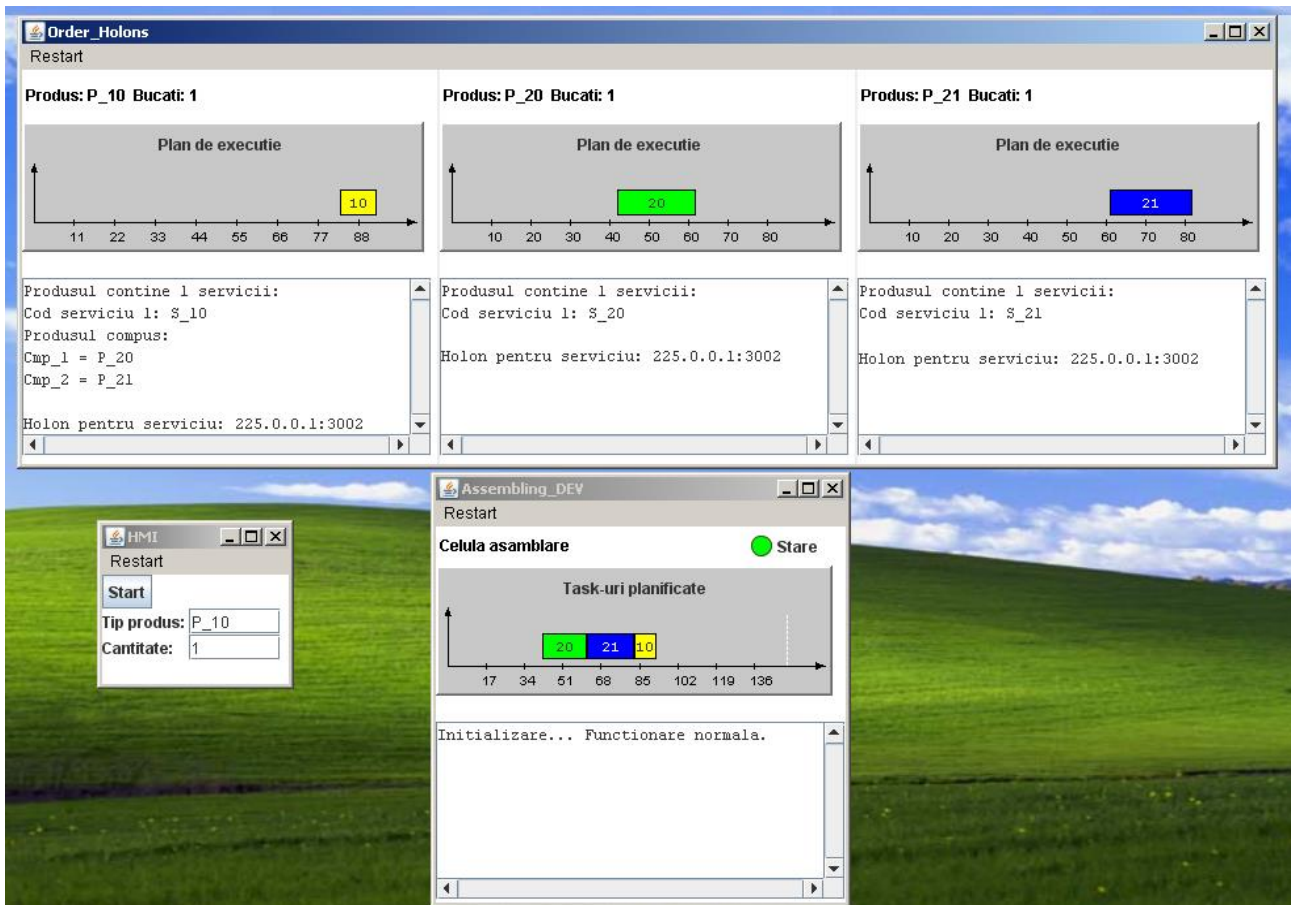

Fig. 19. Exemplificare a interfeţelor om-maşină ale holonilor comandă şi a holonului celulei

Comparativ cu soluţia clasică, soluţia de control holonică permite reducerea timpilor de reconfigurare a celulei, la schimbarea tipului de produs, şi o mai bună monitorizare a activităţilor atribuite celulei.

În modul clasic, schimbarea tipului de produs ar necesita o configurare manuală a celulei, cu durata $T_a$, pentru „activarea" serviciilor noului produs. Acest timp este redus la zero în cazul soluţiei holonice, alegerea serviciilor care trebuie executate la un moment dat fiind realizată automat, funcţie de planificările deţinute de holonul celulei. Pentru serviciile care nu se găsesc în memoria robotului este necesară însă, atât în modul clasic, cât şi în cel holonic, o încărcare manuală a acestora.

## V. CONCLUZII

Obiectivul principal în realizarea acestui studiu de caz a constat în demonstrarea posibilităţii de adoptare a conceptelor IEC 61499 (alături de tehnologii multi-agent) în modelarea şi implementarea sistemelor de control holonice. În acest sens au fost propuse şi verificate modele şi teorii privind arhitectura holonilor individuali, respectiv formarea de holoni complecşi prin descompunerea task-urilor în subtask-uri.

Contribuţiile sunt în principal de ordin practic, şi constau în:

- Dezvoltarea şi testarea unei soluţii de modelare şi implementare a holonilor cu inteligenţă simplă prin blocuri funcţionale IEC 61499;
    o Dezvoltarea unei infrastructuri software pentru comunicaţia dintre holoni, bazată pe modelul de comunicaţie PUBLISH / SUBSCRIBE;
    o Dezvoltarea şi testarea unei soluţii de

implementare a interfeţelor om-maşină ale holonilor prin blocuri funcţionale. Această soluţie s-a dovedit greoaie în contextul instrumentelor de dezvoltare existente;
- Dezvoltarea şi testarea unui protocol de comunicaţie de nivel înalt, pentru comunicaţia inter-holonică. Împreună cu infrastructura software, acest protocol a permis definirea unei platforme multi-agent, care poate fi privită ca o extindere a funcţionalităţilor cuprinse în specificaţiile standardului IEC 61499.
- Dezvoltarea unei soluţii de integrare a echipamentelor de producţie clasice în sisteme de control holonice;
- Implementarea şi testarea unei soluţii de creare şi eliminare dinamică a holonilor comandă, în/din cadrul unor dispozitive IEC 61499, prin comenzi de management.